\documentclass{moriond}

\def\be{\begin{equation}}
\def\ee{\end{equation}}
\def\bea{\begin{eqnarray}}
\def\eea{\end{eqnarray}}

\newcommand{\Photo}{}

\usepackage{amsmath}
\usepackage{amssymb}
\usepackage{subfigure}

\begin{document}
\vspace*{4cm}
\title{Kaon Decays beyond the Standard Model}

\author{Stefan Schacht}

\address{
Department of Physics and Astronomy, University of Manchester,\\ Manchester M13 9PL, United Kingdom
}

\maketitle\abstracts{
We review a new method in order to determine the parameter $\bar{\eta}$ of the Cabibbo-Kobayashi-Maskawa matrix from $K\rightarrow \mu^+\mu^-$ decays, using interference effects in the time-dependent decay rate. Furthermore, we discuss a new precision relation for the phase-shift of the time-dependent oscillation. The new methodology enables the discovery potential of future time-dependent measurements of $K\rightarrow \mu^+\mu^-$ decays for physics beyond the Standard Model.}

\section{Introduction}

Even over 75 years after the discovery of the kaon in Manchester~\cite{Rochester:1947mi}, kaon physics is an exciting field with many new developments.
On the theory side, there has been a lot of renewed interest in the decay $K\rightarrow \mu^+\mu^-$, see recently Refs.~\cite{DAmbrosio:2017klp,Dery:2021mct,Buras:2021nns,Dery:2021vql,DAmbrosio:2022kvb,Brod:2022khx,Dery:2022yqc}.
On the experimental side, recent developments in rare kaon decays have been the $3.4\sigma$ evidence for $K^+\rightarrow \pi^+\nu\bar{\nu}$ at NA62~\cite{NA62:2021zjw}, an improved upper limit on $K_L\rightarrow \pi^0 \nu\bar{\nu} $ from KOTO~\cite{KOTO:2020prk}, as well as 
new upper limits on $\mathcal{B}(K_S\rightarrow \mu^+\mu^-)$~\cite{LHCb:2020ycd} and $\mathcal{B}(K_{S,L}\rightarrow 2(\mu^+\mu^-))$~\cite{LHCb:2022diq} from LHCb. Furthermore, recently, new ideas for the future of kaon physics at CERN have been brought forward~\cite{HIKE:2022qra}.

In this review, I focus on the recent idea to use the time dependence of $K\rightarrow \mu^+\mu^-$ decays as a probe for new physics~\cite{DAmbrosio:2017klp,Dery:2021mct,Brod:2022khx,Dery:2022yqc}. First experimental studies of this idea have been presented in Ref.~\cite{Marchevski:2023kab}.
The new idea is that we can in principle very cleanly measure $\mathrm{Im}(V_{td}^* V_{ts})$, or equivalently~$\bar{\eta}$, from $K\rightarrow \mu^+\mu^-$. We can do so by employing time-dependent interference effects.
In this way, $K\rightarrow \mu^+\mu^-$ is transformed into a third golden mode~\cite{Dery:2022bhj} along $K^+\rightarrow \pi^+\nu\bar{\nu}$ and $K_L\rightarrow \pi^0 \nu\bar{\nu}$ which are currently measured at NA62~\cite{NA62:2020fhy} and KOTO~\cite{KOTO:2020prk}, respectively.

Our new method includes $f_K$ as the main hadronic uncertainty, so it is theoretically clean, however it includes measuring the time-dependent interference effects, which are experimentally challenging.

In Sec.~\ref{sec:main-idea} we present the main idea, giving us a new handle on the Wolfenstein parameter~$\bar{\eta}$. In Sec.~\ref{sec:phase-shift} we present another precision relation related to the phase shift in the time dependence of $K\rightarrow \mu^+\mu^-$ decays.
Constraints on new physics are briefly discussed in Sec.~\ref{sec:new-physics}, before we conclude in Sec.~\ref{sec:conclusion}.

\section{Separating Long- and Short-Distance Physics in $K\rightarrow \mu^+\mu^-$ \label{sec:main-idea}}

One of the long-term goals of the physics program of rare kaon decays is to determine the unitarity triangle purely with kaon decays.
This gives a crucial intergenerational consistency check of the Standard Model (SM) and new ways to probe for new physics.
A key issue for this goal is the identification of observables with a theoretically clean sensitivity to CKM matrix elements.
In order to achieve that, we need methods with a theory error on the hadronic physics at the order of $\sim 1\%$.
In $K\rightarrow \mu^+\mu^-$ we are currently not able to achieve such a theory precision for the long-distance (LD) effects. 
The question is therefore how to extract the short-distance (SD) physics from $K\rightarrow \mu^+\mu^-$ measurements.

In principle, the measurement of the branching ratio $\mathcal{B}(K_S\rightarrow (\mu\mu)_{l=0})$, 
where the index $l=0$ indicates the angular momentum of the muons in the final state,
provides such a clean probe for SD physics. The reason is that, as a transition from a CP-even to a CP-odd state, $\vert A(K_S\rightarrow (\mu\mu)_{l=0})\vert$ is a CP-violating amplitude. As such it has to a very good approximation no contributions from long-distance (LD) physics.
However, in practice final state muons with specific angular momentum $(\mu\mu)_{l=0}$ and $(\mu\mu)_{l=1}$ are not available to us because in 
the decay rate we measure their incoherent sum. 
The key question is therefore how to access $\mathcal{B}(K_S\rightarrow (\mu\mu)_{l=0})$.

At this point it is instructive to take a step back and look at the anatomy of long- and short-distance physics in 
$K\rightarrow \mu^+\mu^-$ in general:~\cite{Dery:2021mct}
\begin{itemize}
\item CP-conserving amplitudes: both SD and LD contributions.\\
	CP-odd $\rightarrow$ CP-odd: $\vert A(K_L\rightarrow (\mu\mu)_{l=0})\vert$\,, \\ 
	CP-even $\rightarrow$ CP-even: $\vert A(K_S\rightarrow (\mu\mu)_{l=1})\vert$.
\item CP-violating amplitudes: only SD contributions.\\
	CP-even $\rightarrow$ CP-odd: $\vert A(K_S\rightarrow (\mu\mu)_{l=0})\vert$\,, \\ 
 	CP-odd $\rightarrow$ CP-even: $\vert A(K_L\rightarrow (\mu\mu)_{l=1})\vert$.
\item Relative phases: both SD and LD contributions.\\
	$\varphi_0 \equiv \mathrm{arg}\left(\mathcal{A}^*(K_S\rightarrow (\mu\mu)_{l=0}) \mathcal{A}(K_L\rightarrow (\mu\mu)_{l=0})\right)$\,, \\
	$\varphi_1 \equiv \mathrm{arg}\left(\mathcal{A}^*(K_S\rightarrow (\mu\mu)_{l=1}) \mathcal{A}(K_L\rightarrow (\mu\mu)_{l=1})\right)$.
\end{itemize}
In our discussion we neglect the small CP violation from mixing, \emph{i.e.},~we take the limit $\varepsilon_K=0$, which can however also be incorporated into the analysis as shown in Ref.~\cite{Brod:2022khx}. 
In the SM, the SD operator does not generate a $(\mu\mu)_{l=1}$ state due to CPT, see,~\emph{e.g.},~the appendix of Ref.~\cite{Dery:2021mct}, and therefore $\vert A(K_L\rightarrow (\mu\mu)_{l=1})\vert = 0$ and $\varphi_1=0$. Because of that, we are left in total with four theory parameters, one of which, namely $\vert A(K_S\rightarrow (\mu\mu)_{l=0} )\vert$, is purely due to SD physics. 
As said above, we can cleanly calculate it in the SM~\cite{Brod:2022khx,Isidori:2003ts,GomezDumm:1998gw,Inami:1980fz} 
\begin{align}
\mathcal{B}(K_S\rightarrow (\mu\mu)_{l=0} ) &= 1.7\cdot 10^{-13}\times \left(\frac{A^2\lambda^5 \bar{\eta}}{1.3\cdot 10^{-4} } \right)\,. \label{eq:SM-prediction}
\end{align}
The hadronic uncertainties from $f_K$~\cite{FlavourLatticeAveragingGroupFLAG:2021npn} as well as from higher-order QCD/EW corrections~\cite{Brod:2022khx} in the prefactor in Eq.~(\ref{eq:SM-prediction}) are at the level of $\sim 1\%$. 
The observable $\mathcal{B}(K_S\rightarrow (\mu\mu)_{l=0} )$ therefore opens the way to a theoretically clean extraction of $\bar{\eta}$, and importantly we can also calculate it cleanly in models beyond the Standard Model (BSM).

Now, the solution to the problem how to access $\mathcal{B}(K_S\rightarrow (\mu\mu)_{l=0} )$ experimentally is as follows.
It consists in measuring the time dependence of $K\rightarrow \mu^+\mu^-$, which can be written as~\cite{ParticleDataGroup:2022pth} 
\begin{align}
\frac{d\Gamma}{dt} \propto C_L e^{-\Gamma_L t} + C_S e^{-\Gamma_S t} + 2 C_{\mathrm{\mathrm{Int.}}} \cos(\Delta m_K t - \varphi_0) e^{-\frac{\Gamma_L+\Gamma_S}{2} t}\,,  \label{eq:time-dependence}
\end{align}
where $C_{L,S}$ are related to the $K_{L,S}$ decay rates, respectively, $C_{\mathrm{Int.}}$ is due to the interference between $K_L$ and $K_S$ decays and $\varphi_0$ is the phase shift of the oscillation. Furthermore, $\Gamma \equiv (\Gamma_S+\Gamma_L)/2$ and $\Delta m$ is the kaon mass difference.
For the example of a pure $K^0$ beam, the four experimental observables in Eq.~(\ref{eq:time-dependence}) can be expressed as follows by the four theory parameters:
\begin{align}
C_L &= \vert A(K_L)_{l=0}\vert^2\,, & 
C_S &= \vert A(K_S)_{l=0}\vert^2 + \beta_{\mu}^2 \vert A(K_S)_{l=1}\vert^2\,,\\
C_{\mathrm{Int.}} &= \vert A(K_S)_{l=0}\vert\, \vert A(K_L)_{l=0}\vert\,, &
\varphi_0 &= \mathrm{arg}\left(A(K_S)_{l=0}^* A(K_L)_{l=0}\right)\,, 
\end{align}
where $\beta_{\mu} = \sqrt{1 - 4 m_{\mu}^2/m_{K^0}^2}$.
Consequently, we can completely solve the system and obtain~\cite{Dery:2021mct}
\begin{align}
\mathcal{B}(K_S\rightarrow (\mu^+\mu^-)_{l=0}) &= \mathcal{B}(K_L\rightarrow \mu^+\mu^- )\cdot\frac{\tau_S}{\tau_L}\cdot\frac{C_{\mathrm{Int}}^2}{C_L^2}\,. \label{eq:solution}
\end{align}
The comparison of Eq.~(\ref{eq:solution}) as extracted from future data with the SM result Eq.~(\ref{eq:SM-prediction}) enables the extraction of $\bar{\eta}$.
This implies that it is crucial to obtain the interference terms in the time dependence Eq.~(\ref{eq:time-dependence}).
We note that Eq.~(\ref{eq:solution}) is only valid in the limit of a pure $K^0$ beam. In a realistic scenario of a mixed $K^0$/$\bar{K}^0$ beam, a dilution factor reduces the sensitivity by 
a respective amount, see Ref.~\cite{Dery:2021mct} for details.

\section{A Precision Prediction for the Phase Shift \label{sec:phase-shift} }

\begin{figure}[t]
\begin{center}
\subfigure[]{\includegraphics[width=0.3\textwidth]{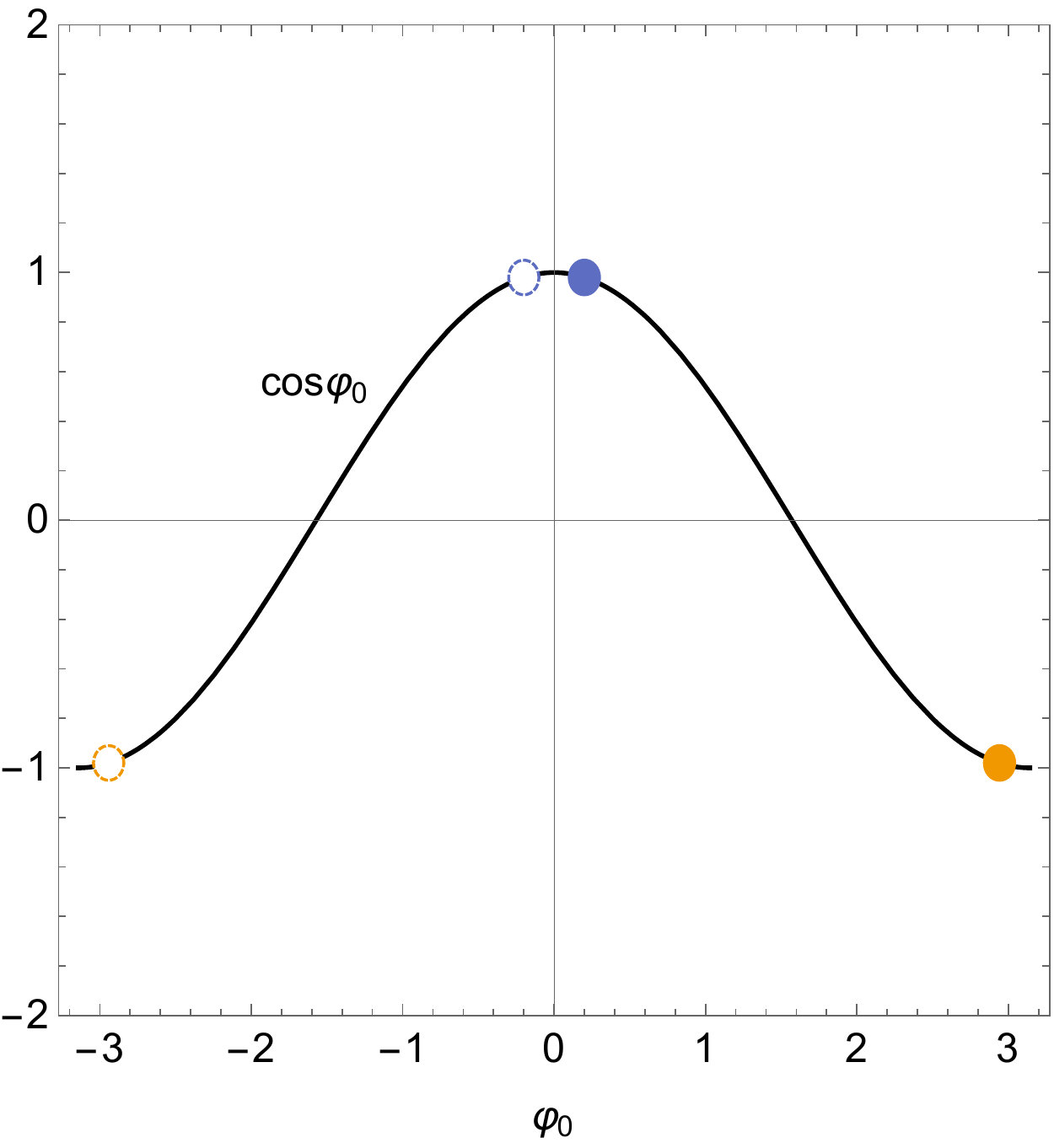}}
\subfigure[]{\includegraphics[width=0.32\textwidth]{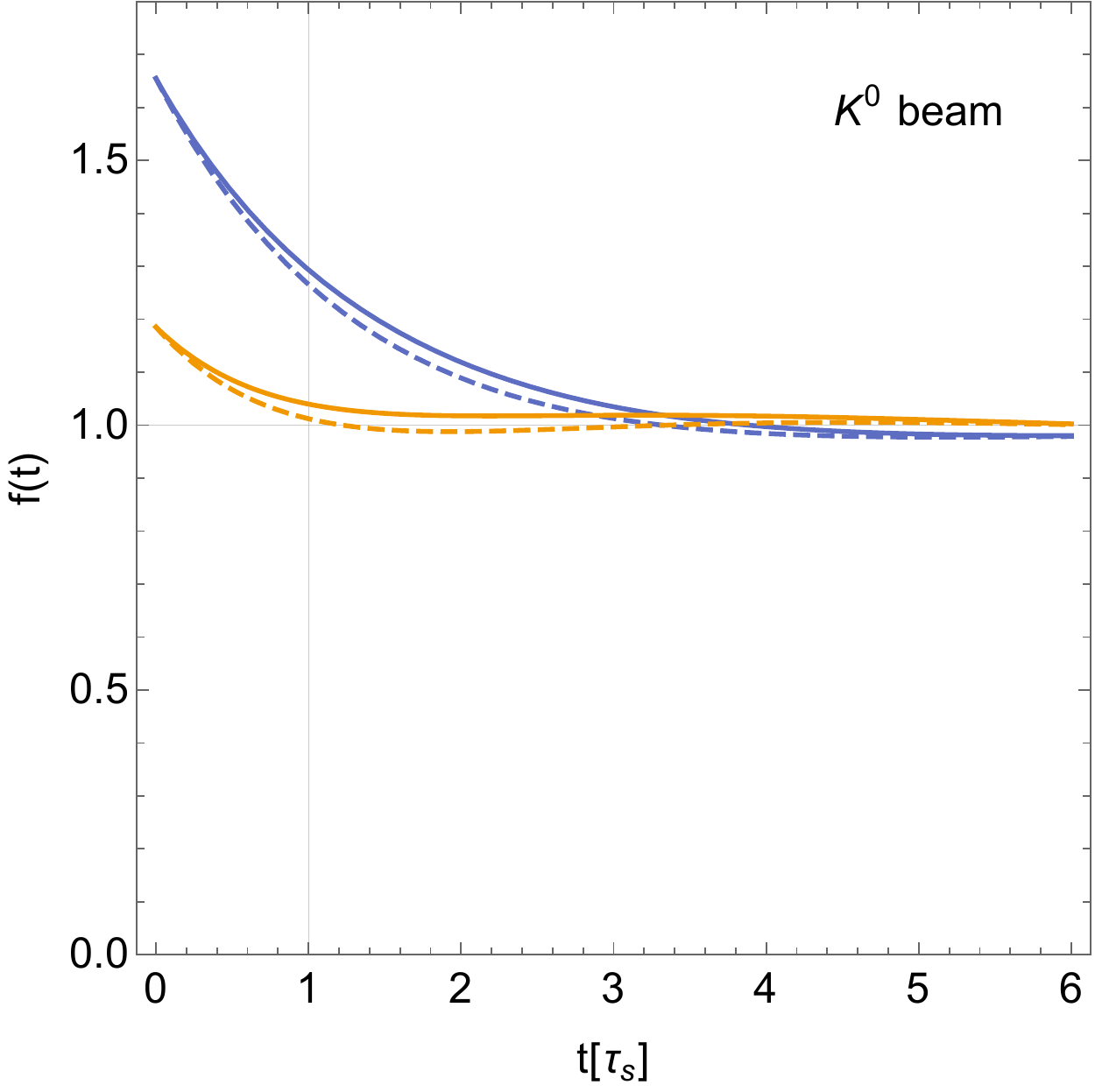}}
\subfigure[]{\includegraphics[width=0.32\textwidth]{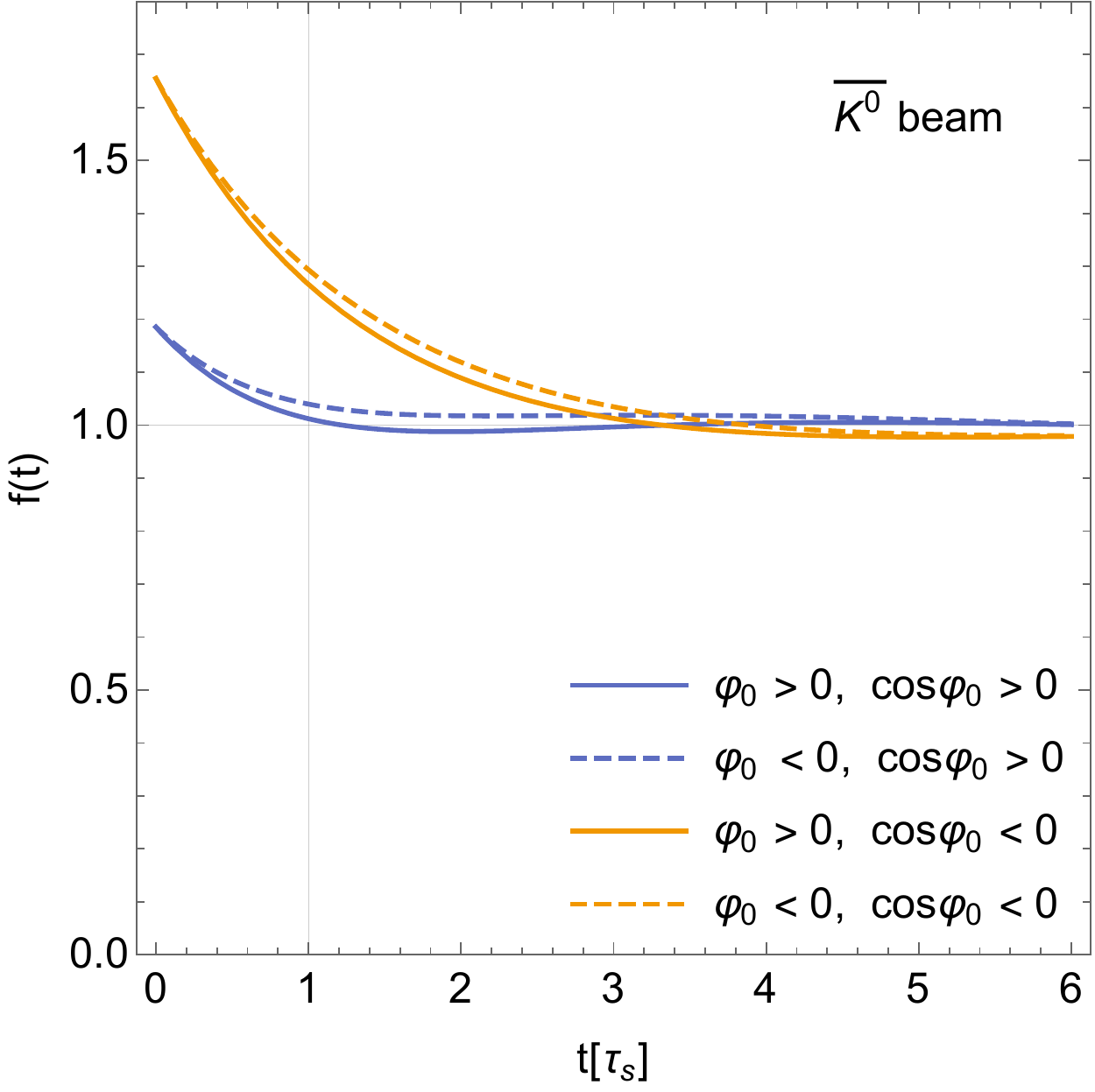}}
\caption[]{The four solutions of Eq.~(\ref{eq:phase-shift}) for $\varphi_0$~(a) and their implications for the time dependence of a pure $K^0$~(b) or $\bar{K}^0$ beam~(c). Figure reproduced from Ref.~\cite{Dery:2022yqc}.}
\label{fig:plot}
\end{center}
\end{figure}

Apart from Eq.~(\ref{eq:solution}), there is one more precision relation for an observable of the time dependence of 
$K\rightarrow \mu^+\mu^-$ that can be used for a SM test. 
The phase shift $\varphi_0$ in Eq.~(\ref{eq:time-dependence}) is related to a known ratio of branching ratios as~\cite{Dery:2022yqc} 
\begin{align}
\cos^2\varphi_0 &= C^2_{\mathrm{QED}} \cdot \frac{\mathcal{B}(K_L\rightarrow \gamma\gamma)}{\mathcal{B}(K_L\rightarrow \mu^+\mu^-) }\,, \label{eq:phase-prediction}
\end{align} 
where $C_{\mathrm{QED}}^2$ is a known QED factor. Eq.~(\ref{eq:phase-prediction}) implies the model-independent prediction~\cite{Dery:2022yqc}
\begin{align}
\cos^2\varphi_0 &= 0.96\pm 0.02_{\mathrm{exp}}\pm 0.02_{\mathrm{th}}\,, \label{eq:phase-shift} 
\end{align}
that can be tested with future measurements of the time dependence of $K\rightarrow\mu^+\mu^-$.
In Eq.~(\ref{eq:phase-shift}), the experimental error comes from the one of the involved branching ratios.
The theory error comes from higher order QED corrections and from additional intermediate on-shell contributions 
that give a small correction to the dominant two-photon contribution~\cite{Martin:1970ai}.
Eq.~(\ref{eq:phase-shift}) has actually four model-independent solutions for $\varphi_0$, each of which implies a different time evolution. We show 
these solutions in Fig.~\ref{fig:plot}.

\section{Constraints on New Physics \label{sec:new-physics}}

Although no measurement of the time dependence of $K\rightarrow \mu^+\mu^-$ is available, the LHCb constraint on the branching ratio~\cite{LHCb:2020ycd}
\begin{align}
\mathcal{B}(K_S\rightarrow \mu^+\mu^- )  < 2.1\cdot 10^{-10} \label{eq:KS-bound}
\end{align} 
constrains relevant parameter space of new physics models already now, as is worked out in detail in Ref.~\cite{Dery:2021vql}.
Therein, in order to be conservative, the bound Eq.~(\ref{eq:KS-bound}) is interpreted as a bound on $\mathcal{B}(K_S\rightarrow \mu^+\mu^-)_{l=0}$, resulting in a lot of room 
for BSM physics to be tested~\cite{Dery:2021vql} 
\begin{align}
\frac{\mathcal{B}(K_S\rightarrow \mu^+\mu^-)_{l=0}^{\mathrm{SM}}}{\mathcal{B}(K_S\rightarrow \mu^+\mu^-)_{l=0}} \leq 1280\,. \label{eq:bound}
\end{align}
However, scalar leptoquark or two-Higgs doublet models that saturate the bound Eq.~(\ref{eq:bound}) can be constructed, at the same time being 
consistent with existing constraints, see Ref.~\cite{Dery:2021vql} for details. 
The decays $K\rightarrow \mu^+\mu^-$ and $K_L\rightarrow \pi^0 \nu \bar{\nu}$, although sensitive to the same CKM matrix element combination in the SM,
are sensitive to different new physics operators in BSM models~\cite{Dery:2021vql}.
Future updated bounds of the constraint Eq.~(\ref{eq:KS-bound}) are important to probe the parameter space of BSM models further.

\section{Conclusion \label{sec:conclusion} }

The time dependence of $K\rightarrow \mu^+\mu^-$ gives two independent SM tests. The coefficient of the interference term of the time-dependent decay rate 
is sensitive to $\mathcal{B}(K_S\rightarrow \mu^+\mu^-)_{l=0}$, which in the SM is proportional to the Wolfenstein parameter $\bar{\eta}$.
The second SM test is given by a precision relation of the oscillation phase shift, which is predicted model-independently up to a four-fold ambiguity.
The leptonic kaon decay mode $K\rightarrow \mu^+\mu^-$ turns out to be theoretically clean and experimentally challenging, similar in that respect to the related decay modes
$K^+\rightarrow \pi^+\nu\bar{\nu}$ and $K_L\rightarrow \pi^0 \nu\bar{\nu}$.

\section*{Acknowledgments}

I thank my collaborators Avital Dery, Mitrajyoti Ghosh, Yuval Grossman and Teppei Kitahara, together with whom the works reviewed above have been carried out. 
S.S.~is supported by a Stephen Hawking Fellowship from UKRI under reference EP/T01623X/1 and the Lancaster-Manchester-Sheffield Consortium for Fundamental Physics, under STFC research grant ST/T001038/1. 

\section*{References}
\begin{scriptsize}
\bibliography{schacht}

\begin{thebibliography}{10}

\bibitem{Rochester:1947mi}
G.~D. Rochester and C.~C. Butler.
\newblock {Evidence for the Existence of New Unstable Elementary Particles}.
\newblock {\em Nature}, 160:855--857, 1947.

\bibitem{DAmbrosio:2017klp}
Giancarlo D'Ambrosio and Teppei Kitahara.
\newblock {Direct $CP$ Violation in $K \to \mu^+ \mu^-$}.
\newblock {\em Phys. Rev. Lett.}, 119(20):201802, 2017.

\bibitem{Dery:2021mct}
Avital Dery, Mitrajyoti Ghosh, Yuval Grossman, and Stefan Schacht.
\newblock {$K \to {\mu}^{+} \mu^-$ as a clean probe of short-distance physics}.
\newblock {\em JHEP}, 07:103, 2021.

\bibitem{Buras:2021nns}
Andrzej~J. Buras and Elena Venturini.
\newblock {Searching for New Physics in Rare $K$ and $B$ Decays without
  $|V_{cb}|$ and $|V_{ub}|$ Uncertainties}.
\newblock {\em Acta Phys. Polon. B}, 53(6):A1, 9 2021.

\bibitem{Dery:2021vql}
Avital Dery and Mitrajyoti Ghosh.
\newblock {$K\rightarrow \mu^+\mu^-$ beyond the standard model}.
\newblock {\em JHEP}, 03:048, 2022.

\bibitem{DAmbrosio:2022kvb}
G.~D'Ambrosio, A.~M. Iyer, F.~Mahmoudi, and S.~Neshatpour.
\newblock {Anatomy of kaon decays and prospects for lepton flavour universality
  violation}.
\newblock {\em JHEP}, 09:148, 2022.

\bibitem{Brod:2022khx}
Joachim Brod and Emmanuel Stamou.
\newblock {Impact of indirect CP violation on Br$(K_S \to
  \mu^+\mu^-)_{\ell=0}$}.
\newblock 9 2022.

\bibitem{Dery:2022yqc}
Avital Dery, Mitrajyoti Ghosh, Yuval Grossman, Teppei Kitahara, and Stefan
  Schacht.
\newblock {A Precision Relation between $\Gamma(K\to\mu^+\mu^-)(t)$ and ${\cal
  B}(K_L\to\mu^+\mu^-)/{\cal B}(K_L\to\gamma\gamma)$}.
\newblock {\em JHEP}, 03:014, 2023.

\bibitem{NA62:2021zjw}
Eduardo Cortina~Gil et~al.
\newblock {Measurement of the very rare K$^{+}$\textrightarrow{}$ {\pi}^{+}\nu
  \overline{\nu} $ decay}.
\newblock {\em JHEP}, 06:093, 2021.

\bibitem{KOTO:2020prk}
J.~K. Ahn et~al.
\newblock {Study of the $K_L \to \pi^0 \nu \bar \nu$ Decay at the J-PARC KOTO
  Experiment}.
\newblock {\em Phys. Rev. Lett.}, 126(12):121801, 2021.

\bibitem{LHCb:2020ycd}
Roel Aaij et~al.
\newblock {Constraints on the $K^0_S \rightarrow \mu^+ \mu^-$ Branching
  Fraction}.
\newblock {\em Phys. Rev. Lett.}, 125(23):231801, 2020.

\bibitem{LHCb:2022diq}
{Search for $K^0_{\mathrm{S(L)}}\to\mu^+\mu^-\mu^+\mu^-$ decays at LHCb}.
\newblock 12 2022.

\bibitem{HIKE:2022qra}
E.~Cortina~Gil et~al.
\newblock {HIKE, High Intensity Kaon Experiments at the CERN SPS}.
\newblock 11 2022.

\bibitem{Marchevski:2023kab}
Radoslav Marchevski.
\newblock {First thought on a high-intensity $K_S$ experiment}.
\newblock In {\em {International Conference on Kaon Physics 2022}}, 1 2023.

\bibitem{Dery:2022bhj}
Avital Dery.
\newblock {K \textrightarrow{} \ensuremath{\mu}+\ensuremath{\mu}\ensuremath{-}
  as a third kaon golden mode}.
\newblock {\em J. Phys. Conf. Ser.}, 2446(1):012034, 2023.

\bibitem{NA62:2020fhy}
Eduardo Cortina~Gil et~al.
\newblock {An investigation of the very rare $ {K}^{+}\to {\pi}^{+}\nu
  \overline{\nu} $ decay}.
\newblock {\em JHEP}, 11:042, 2020.

\bibitem{Isidori:2003ts}
Gino Isidori and Rene Unterdorfer.
\newblock {On the short distance constraints from $K_{L,S} \to \mu^+\mu^- $}.
\newblock {\em JHEP}, 01:009, 2004.

\bibitem{GomezDumm:1998gw}
D.~Gomez~Dumm and A.~Pich.
\newblock {Long distance contributions to the $K_L \to \mu^+\mu^- $ decay
  width}.
\newblock {\em Phys. Rev. Lett.}, 80:4633--4636, 1998.

\bibitem{Inami:1980fz}
T.~Inami and C.~S. Lim.
\newblock {Effects of Superheavy Quarks and Leptons in Low-Energy Weak
  Processes k(L) ---\ensuremath{>} mu anti-mu, K+ ---\ensuremath{>} pi+
  Neutrino anti-neutrino and K0 \ensuremath{<}---\ensuremath{>} anti-K0}.
\newblock {\em Prog. Theor. Phys.}, 65:297, 1981.
\newblock [Erratum: Prog.Theor.Phys. 65, 1772 (1981)].

\bibitem{FlavourLatticeAveragingGroupFLAG:2021npn}
Y.~Aoki et~al.
\newblock {FLAG Review 2021}.
\newblock {\em Eur. Phys. J. C}, 82(10):869, 2022.

\bibitem{ParticleDataGroup:2022pth}
R.~L. Workman et~al.
\newblock {Review of Particle Physics}.
\newblock {\em PTEP}, 2022:083C01, 2022.

\bibitem{Martin:1970ai}
B.~R. Martin, E.~De~Rafael, and J.~Smith.
\newblock {Neutral kaon decays into lepton pairs}.
\newblock {\em Phys. Rev. D}, 2:179--200, 1970.

\end{thebibliography}
\end{scriptsize}
\end{document}